# Atomistic Study of the Electronic Contact Resistivity Between the Half-Heusler Alloys (HfCoSb, HfZrCoSb, HfZrNiSn) and the Metal Ag

Yuping He, François Léonard and Catalin D. Spataru*

*Sandia National Laboratories*, *Livermore, California, 94551, USA*
*E-mail: cdspata@sandia.gov

## Abstract

Half-Heusler(HH) alloys have shown promising thermoelectric properties in the medium and high temperature range. To harness these material properties for thermoelectric applications, it is important to realize electrical contacts with low electrical contact resistivity. However, little is known about the detailed structural and electronic properties of such contacts, and the expected values of contact resistivity. Here, we employ atomistic *ab initio* calculations to study electrical contacts in a subclass of HH alloys consisting of the compounds HfCoSb, HfZrCoSb, and HfZrNiSn. By using Ag as a prototypical metal, we show that the termination of the HH material critically determines the presence or absence of strong deformations at the interface. Our study includes contacts to doped materials, and the results indicate that the p-type materials generally form ohmic contacts while the n-type materials have a small Schottky barrier. We calculate the temperature dependence of the contact resistivity in the low to medium temperature range and provide quantitative values that set lower limits for these systems.

# I. Introduction

Thermoelectric (TE) devices convert heat into electricity, or vice versa, without any moving parts and are therefore attractive for energy harvesting and electronic cooling. A typical thermoelectric module is an array of n-type and p-type semiconducting materials connected electrically in series and thermally in parallel. The efficiency ($\eta$) of thermoelectric devices is given by [1]

$$\eta = \eta_{max}\left(\frac{\sqrt{1+ZT}-1}{\sqrt{1+ZT}+T_C/T_H}\right) \qquad (1)$$

where $\eta_{max}=(T_H-T_C)/T_H$ is the Carnot efficiency, $T_H$ and $T_C$ are the hot and cold side temperatures, and $ZT$ is a dimensionless figure of merit that characterizes the TE performance of materials. For the efficiency of a TE device, ZT is an average of the ZT values for the n-type and p-type TE materials. The material ZT is given by $(ZT)_{mat} = S^2\sigma T/\kappa$, where $\sigma$ is the electrical conductivity, S the Seebeck coefficient, and $\kappa$ the thermal conductivity. Equation 1 is valid in the ideal situation where the thermal and electrical contact resistivity at the hot and cold ends are negligible [2]. In reality, the finite electrical and thermal contact resistivities lead to both temperature and voltage drops at the hot and cold ends, hence reduce the overall device efficiency, which is proportional to the effective figure of merit $(ZT)_{eff} = ZT * [1 + \frac{2\rho_C}{\rho_m L_m}]^{-1}$, where ZT is the device level figure of merit without contact effects, $\rho_C$ is the contact resistivity, $\rho_m$ and $L_m$ are the resistivity and length of the p-type and n-type TE materials for the legs in the TE module. Thus, for optimal device performance, the contact resistivity should be as small as possible; a rule of thumb is that a good module performance requires $\rho_C<\rho*L/10$.

Half-Heusler (HH) alloys have recently attracted a great deal of attention as promising thermoelectric materials [3-5] for medium and high temperatures (i.e. in the range of 500 to 1000 K) due to their high chemical stability and small lattice thermal conductivity, which is achieved by creating nanostructures, atomic mass differences, and multiple alloys[6-8]. Within the family of HH alloys, the p-type $Hf_{0.5}Zr_{0.5}CoSb_{0.8}Sn_{0.2}$ material and the n-type $Hf_{0.75}Zr_{0.25}NiSn_{0.99}Sb_{0.01}$ material have shown large ZT values of 0.8 and 1, respectively[9]. In addition, these materials are relatively low-cost and environmentally friendly hence desired for green energy generation.

To harness these promising material properties for thermoelectric applications requires addressing a number of challenges. One such challenge is to identify appropriate metal electrodes that form good electrical and thermal contacts at elevated operating temperatures[10-13]. Several metal electrodes have been explored, including Ag[14], Cu[5], stainless steel[15] and Ti[16]. Among these candidates, Ag is a promising option owing to its stability at elevated temperature, and high electrical and thermal conductivity. Recently Ngan *et al.* [14] showed that by using fast hot pressing technique to join HH with metal Ag, a contact electrical resistivity of $\sim 5 \times 10^{-5}$ $\Omega cm^2$ can be achieved for temperatures up to 450 °C. Still, order-of-magnitude lower contact resistivity is desired for the ideal performance of compact, high-power-density TE generators based on HH alloys with intrinsic resistivity $\rho \sim 1 m\Omega cm$ [17] and with typical size $L \sim 1 mm$.

HH materials have been extensively studied both theoretically [4,18,19] and experimentally[6,9,19,20], to establish their basic electronic properties as well as to improve their thermoelectric (TE) properties for thermal energy conversion applications[19]. The investigation of contacts between HH materials and metal electrodes is however very rare. One of the challenges in developing high efficiency TE generators based on HHs is the absence of a fundamental knowledge of physical and chemical properties of interfaces between HH materials and metal contacts. *Ab initio* simulation is a powerful tool to explore material properties at the atomic scale including interfaces and surfaces. Nevertheless, *ab initio* calculations of the properties of electrical contacts between HH materials and metals have not been theoretically reported. In this rapid communication, we present an atomistic study of electrical contact resistivity between n- and p-type HHs and metal silver (Ag), using a combination of *ab initio* simulations and macroscopic modeling, to explore the effects of interfacial structure, chemical composition and doping on the electronic contact resistivity. Interfacial band bending and electrical contact resistivity have been calculated for the different interfacial structures of HH and Ag. We find that the termination of the HH plays an important role in determining the structure of the interface, which in turn induces different band bending at the interfaces, whereas far away from the interfaces, the band bending is mainly dominated by the electronic doping. Our results indicate that the detailed atomic and electronic structure of the interface plays a key role in determining the band alignment and ultimately the contact resistivity. This adds to the growing body of work on metal/semiconductor interfaces[21,22] that demonstrates the need to carefully consider atomistic effects when calculating Schottky barriers, and more generally, band alignment. Indeed, the possibility of *ab*

*initio* computations on large unit cells now allows a more robust calculation of complex epitaxial interfaces, including disorder. This gives a more refined picture that reveals unexpected interfacial effects, such as the role of chemical dipoles discussed in this manuscript.

**Table I** Generated Half-Heusler simulation cells with different characteristics (i.e. n- or p-type doping). The optimized lattice constants ($a_x$, $a_y$ and $a_z$) and bandgaps obtained by *ab initio* DFT-GGA. The carrier concentration is estimated by assuming that one dopant donates one carrier. The Schottky barrier is obtained from the *ab initio* based modeling.

| Half-Heusler | Characteristic | Unit Cell | $a_x$ /Å | $a_y$ /Å | $a_z$ /Å | Band Gap /eV | Carrier Concentration /cm$^{-3}$ | Schottky Barrier /eV |
|---|---|---|---|---|---|---|---|---|
| HfCoSb | Intrinsic Semi. | 1x1x1 | 6.0661 | 6.0661 | 6.0661 | 1.12 | | |
| $Hf_{0.5}Zr_{0.5}CoSb$ | Intrinsic Semi. | 1x1x1 | 6.0881 | 6.0881 | 6.0891 | 1.05 | | |
| $HfCoSn_{0.25}Sb_{0.75}$ | p-type | 1x1x1 | 6.0815 | 6.0815 | 6.0815 | 1.03 | $4.4 \times 10^{21}$ | |
| $Hf_{0.5}Zr_{0.5}CoSn_{0.25}Sb_{0.75}$ | p-type | 1x1x1 | 6.1011 | 6.1011 | 6.1064 | 1.01 | $4.4 \times 10^{21}$ | ohmic |
| $Hf_{0.625}Zr_{0.375}NiSn$ | Intrinsic Semi. | 1x1x2 | 6.1332 | 6.1332 | 12.2657 | 0.41 | | |
| $Hf_{0.625}Zr_{0.375}NiSn_{0.875}Sb_{0.125}$ | n-type | 1x1x2 | 6.1339 | 6.1323 | 12.2579 | 0.41 | $2.2 \times 10^{21}$ | |
| $Hf_{0.625}Zr_{0.375}NiSn_{0.99}Sb_{0.01}$ | n-type | 2x2x8 | 12.2666 | 12.2668 | 49.0681 | 0.41 | $1.3 \times 10^{20}$ | 0.1/0.05 |

## II. Methods

A rigid shift of the local electrostatic potential is used to calculate the contact resistivity within an effective mass model. A good agreement between the effective mass and full *ab initio* approaches has been demonstrated for other semiconductor contacts[23]. We have recently applied this approach[24] to calculate the resistivity of a metal contact to a $(Sb,Bi)_2Te_3$ superlattice, and the obtained results are in very good agreement with experiments.

*Ab initio quantum mechanical calculations*

We carried out *ab initio* calculations of the electronic structure of a series of HH systems belonging to the subclasses HfCoSb, $Hf_{0.5}Zr_{0.5}CoSb$ and $Hf_{0.625}Zr_{0.375}NiSn$. We considered the intrinsic materials as well as those that are p-type ($HfCoSn_{0.25}Sb_{0.75}$ and $Hf_{0.5}Zr_{0.5}CoSn_{0.25}Sb_{0.75}$) and n-type ($Hf_{0.625}Zr_{0.375}NiSn_{0.875}Sb_{0.125}$ and $Hf_{0.625}Zr_{0.375}NiSn_{0.99}Sb_{0.01}$). As shown in Table I, we started from the intrinsic semiconducting HH with a simple cubic crystalline structures, i.e. $Hf_4Co_4Sb_4$ for p-type and $Hf_4Ni_4Sn_4$ for n-type. We then randomly substituted the appropriate species one at a time to form a HH compound with the similar chemical stoichiometry as the experimental samples ($Hf_{0.5}Zr_{0.5}CoSn_{0.25}Sb_{0.75}$ for p-type and $Hf_{0.625}Zr_{0.375}NiSn_{0.99}Sb_{0.01}$ for n-type [14]). For the p-type HH, the unit cell (1x1x1 with 12 atoms) is large enough to construct the system with the hole concentration same as experimental value ($4.4\times10^{21}/cm^{-3}$) by randomly replacing two Hf with Zr, and one Sb with Sn. For n-type HH, in order to form the chemical stoichiometry similar to experimental samples, we tuned the structures in two steps. Firstly, we replicated the unit cell (1x1x1 with 12 atoms) to 1x1x2 with 24 atoms (i.e. $Hf_8Ni_8Sn_8$). We then randomly replaced three Hf by Zr, which yields a chemical formula of $Hf_{0.625}Zr_{0.375}NiSn$. Secondly, we replicated the unit cell of $Hf_{0.625}Zr_{0.375}NiSn$ (1x1x2 with 24 atoms) to 2x2x8 with 384 atoms. We then randomly replaced one Sn by Sb to form n-type HH with the electron concentration similar to the experimental value ($1.3\times10^{20}/cm^{-3}$), leading to the chemical formula of $Hf_{0.625}Zr_{0.375}NiSn_{0.99}Sb_{0.01}$."

The calculations were performed using VASP[25], an *ab initio* simulation package based on Density Functional Theory (DFT). The exchange and correlation terms were treated using the generalized gradient approximation (GGA) with PW91 functional[26]. The projector augmented wave (PAW) pseudo-potentials[27] were chosen to describe core electrons. The valence electron configurations of $5p^66s^25d^2$, $3d^84s^1$, $5s^25p^3$, $5s^25p^2$, $4s^24p^65s^24d^2$ and $4s^23d^8$ were used for Hf, Co, Sb, Sn, Zr and Ni atoms, respectively. A kinetic energy cutoff of 450 eV was used for the expansion of the plane-wave basis set for all systems. K-point grids of 4x4x4, 4x4x2, and 2x2x1 were used to sample the Brillouin zones of the systems of size 1x1x1, 1x1x2 and 2x2x8, the numbers representing the number of repeated unit cells (i.e. simple cubic) in the X, Y and Z directions. We started from the simple HH system of HfCoSb, and then introduced new species one at a time in order to obtain the desired doped-HH systems following the experimentally synthesized structures[14]. The number of the new species are chosen to yield the appropriate electron and hole concentrations. The atomic structures of all systems were optimized with respect to both volume and atomic positions since the optimized lattice constants are sensitive to the chemical compositions and atomic positions (See Table I).

To simulate the interfacial structures between the HH and Ag, we created a slab composed of a HH with a thickness of 5.5 nm and Ag with a thickness of 3.3nm. In this work we considered that the direction perpendicular to the interface corresponds to the [001] direction of the cubic cell, but other directions are also possible[28]. The length of the simulation box along the direction perpendicular to the interface was then enlarged with 4.4 nm of vacuum to avoid image interactions. In the directions parallel to the interface, we employed periodic boundary conditions with a supercell of 1.20x1.20 nm² and 1.24x1.24 nm² for HH and Ag, respectively, which yields a strain of ~2% in these directions. Three possible interfacial structures for the p-type doped HH were simulated, i.e. HfSb/Ag, ZrSbSn/Ag and Co/Ag (the first elements indicate the surface termination of the HH slab contacted by Ag), and two interfacial structures for the n-type doped HH were simulated, HfZrSn/Ag and Ni/Ag, to investigate the effects of interfacial structures and chemical compositions on the interfacial resistivities. All slab systems were optimized with respect to atomic positions again with interatomic force smaller than 0.01 eV/Å. GGA-PW91 and PAW were also used for the simulation of Ag to be consistent with the HH simulation conditions, and the valence electron configuration of Ag atoms was $5s^1 4d^{10}$.

For the doped systems, the carrier concentrations as a function of the Fermi level with respect to the conduction band minimum ($E_{CBM}$) and valence band maximum ($E_{VBM}$) were computed using the standard formula $n = \int_{E_{CBM}}^{+\infty} g(E) f(E) \, dE$ and $p = \int_{-\infty}^{E_{VBM}} g(E) [1 - f(E)] \, dE$ for electron and hole charge carriers, respectively[29]. In these equations, $f(E) = \frac{1}{1 + e^{(E-E_F)/k_B T}}$ is the Fermi Dirac distribution function, and $g(E) = \sum_k W_k (1/w\sqrt{2\pi}) e^{-(E-\varepsilon_k)^2/2w^2}$ is the carrier density of states, where $W_k$ is the weight of k sampling, $E_F$ the Fermi energy level, $k_B$ the Boltzmann constant, $T$ the temperature, and $\varepsilon_k$ is the $k$th electronic state energy. The convergence was checked using a full electronic band structure calculation with different of k-point samplings (i.e. 8x8x4, 24x24x12 and 36x36x18) over a whole Brillouin zone and a Gaussian smearing width of w=0.01 eV.

The density of electronic transport modes was computed using the formula $M(E) = (h/2L) \sum_k W_k |v_k^Z| \delta(E - E_k)$ [30] and $\delta(E - E_k) = (1/w\sqrt{2\pi}) e^{-(E-E_k)^2/2w^2}$, here $h$ is the Planck constant, $L$ is the length of the simulation cell along transport direction (i.e. Z direction), and $v_k^Z = (1/h)(\Delta E_k / \Delta k_Z)$ is the group velocity of charge carriers computed using a finite difference method with a small value of $\Delta k_Z$ along the Z direction. The Gaussian smearing width $w = 0.01$ eV were used, the same as the value in the calculation of carrier concentrations.

*Ab initio band bending*

The band bending in the HH semiconductor was computed by subtracting the planar-averaged electrostatic potentials of the isolated HH slab ($V_{HH}$) and the isolated Ag slab ($V_{Ag}$) from the interfaced slab ($V_{HH-Ag}$), i.e. $V_{inter} = V_{HH-Ag} - V_{HH} - V_{Ag}$. This definition for V results in a smooth function as the ionic potential contribution cancels out exactly. However, it may induce an artificial charge and electrostatic potential on the interface due to the dangling bonds on the surface of the HH and Ag slabs. We have also calculated the band bending by using a macroscopic double-average method[31] along the direction perpendicular to the interface, and found that the artificial charges only affect the electrostatic potential within ~2 Å (or one-bond length) away from the interface. Therefore, we consider the band bending in the HH at distances larger than ~2 Å from the interface. For *ab initio* calculations, an accurate band bending requires a large k-sampling of the Brillouin zone, which is not feasible for large supercell systems. To overcome this issue, we employ a combination of macroscopic modeling and *ab initio* calculations. Hereafter, we call it *ab initio*-based modeling of the band bending. We have applied this approach to the case of n-type doped HH-Ag interface, in four steps:

i) Using *ab initio* simulation, we directly calculate the interfacial electrostatic potential bending *V(z)* for a relatively small (at the mesoscopic scale) HH and Ag interfacial system (1.2 nm x 1.2 nm x 4.9 nm) with ~ 700 atoms, and using the largest k-sampling (2x2x1) computationally feasible given the size of the system.

ii) The Poisson's equation satisfied by V(z) is re-written after integrating out the screening response of the valence bands: $\nabla^2 V(z) = -\rho_0(z) + e[n(z) - p(z) + N_d]/\varepsilon$ which is solved with the definition of free carrier density

$$n(z) = \int_{E_c(z)}^{\infty} D_n(E,z) f(E - E_F) dE \quad (1)$$

$$p(z) = \int_{-\infty}^{E_v(z)} D_p(E,z)[1 - f(E - E_F)] dE \quad (2)$$

Here $\rho_0(z)$ is the charge induced at the interface, $N_d$ is the dopant concentration, $E_c$ and $E_v$ are the CBM and VBM band profiles (rigidly shifted by the local electrostatic potential), and $D_n(E,z)$ and $D_p(E,z)$ are the electron and hole density of states based on effective mass theory:

$$D_n(E,z) = 3 \frac{8\pi\sqrt{2} m_e^{3/2}}{h^3} \sqrt{E - E_c(z)} \quad (3)$$

$$D_p(E,z) = 3\frac{8\pi\sqrt{2}m_h^{3/2}}{h^3}\sqrt{E - E_v(z)} \qquad (4)$$

where the factor of 3 represents the band degeneracy of VBM and CBM.

The equations are discretized on a one-dimensional dense grid and solved self-consistently using linear mixing at each iteration. Poisson's equation satisfies boundary conditions at the two ends of the simulation cell. Far away from the contact we impose two boundary conditions, namely $n(\infty) - p(\infty) + N_d = 0$ and $\partial n(z)/\partial z|_{z=\infty} = \partial p(z)/\partial z|_{z=\infty}$.

The Poisson equation is solved numerically with the input *V(z)* obtained from the *ab initio* calculations for the small interfacial system, to extract the localized (not due to free carriers) interfacial charge $\rho_0$ within 1 nm from the interface, which converges fast with the system size and k-point sampling. $N_d$ is chosen according to the *ab initio* model of n-type doping system, i.e. $1.3\times10^{20}$ cm$^{-3}$, the dielectric constant ε is 22 obtained from our *ab initio* calculations and in good agreement with other first principle results[32], and the effective masses $m_e = 3.03$ m$_o$ and $m_h = 2.74$ m$_o$ are the density of state effective masses for electron and hole, respectively, and obtained by fitting the carrier concentration near the CBM and VBM and taking into account the band degeneracy factor.

iii) We calculate from a*b initio* the converged $m_e = 0.84$ m$_o$ and $m_h = 1.68$ m$_o$ for the intrinsic HH using smaller unit cells with large k-sampling of 36x36x18. Using the converged $m_e$ and $m_h$, and the previously extracted interfacial charge $\rho_0$ we self-consistently solve the Poisson equation again to obtain the converged (w.r.t. k-point sampling) *V(z)*.

## *Macroscopic modeling of contact resistivity*

The contact resistivity is calculated from the current density *J* using

$$\rho_c = \left(\frac{\partial J}{\partial V}\right)^{-1}\bigg|_{V=0} \qquad (5)$$

with [33]

$$\frac{\partial J}{\partial V}\bigg|_{V=0} = -\frac{12\pi m_{DOM} e}{h^3}\int_0^\infty dE \frac{\partial f(E-E_F)}{\partial E}\int_0^E P(E_z)\,dE_z \qquad (6)$$

and

$$P(E_z) = exp\left[-\frac{2\sqrt{6m_{DOM}}}{h}\int \sqrt{eV(z) - E_z}\, dz\right] \quad . \quad (7)$$

The electrostatic potential is calculated from the *ab initio* modeling introduced previously. We use $m_{DOM} \sim 2m_o$ for the density of modes (DOM)[30] effective mass as obtained via a fit near CBM of the DOM (see Fig. 2d). Equation (7) represents the tunneling transmission probability as a function of the component of the carrier energy $E_z$ perpendicular to the interface. Thermionic emission is included in our calculations by setting $P(E_z) = 1$ when the carrier energy is larger than the Schottky barrier height.

## III. Results and Discussion

We first discuss the properties of the bulk HH alloys, which serve as the basis for forming the interfaces with Ag. Table I shows the atomic structure of the *bulk* HH systems investigated in this work for both intrinsic and doped cases. The intrinsic materials have a simple cubic structure with equal lattice constants in all three directions. We construct the doped systems starting from the simple cubic basic unit cell of the intrinsic systems, but after *ab initio* optimization they become pseudo-cubic with slightly different lattice constants in the three directions, depending on the position of dopants. As shown in Figure 1, the bandgap and band structure are not very sensitive to the doping for doping levels of the order of $10^{21}$ cm$^{-3}$. Introducing the dopants rigidly shift the Fermi level into the conduction or valence bands, depending on the substituted atoms. Since Sb has one more valence electron than Sn, replacing Sn by Sb results in a n-type doped system and leads to the Fermi level shifted toward the conduction bands, while replacing Sb by Sn results in a p-type doped system with the Fermi level shifted toward the valence bands (See Figure 1). In this work, we generated n-type (i.e. Hf$_{0.625}$Zr$_{0.375}$NiSn$_{0.99}$Sb$_{0.01}$) and p-type (i.e. Hf$_{0.5}$Zr$_{0.5}$CoSn$_{0.25}$Sb$_{0.75}$) HHs with carrier concentrations of $1.3\times10^{20}$ cm$^{-3}$ and $4.4\times10^{21}$ cm$^{-3}$, respectively, similar to those of the experimental samples[14].

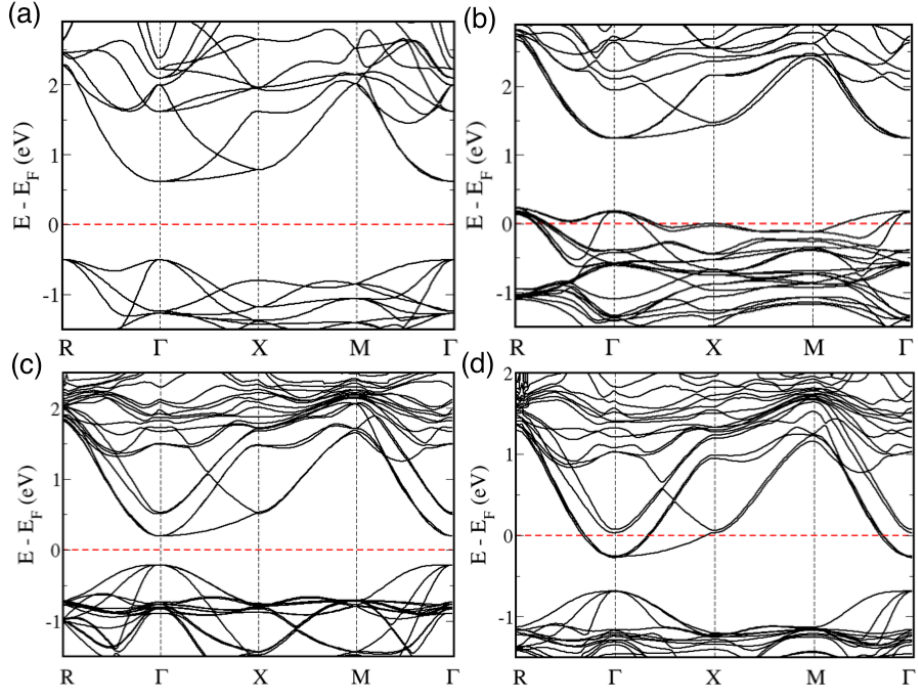

**Figure 1.** Calculated band structures for (a) Intrinsic HfCoSb, (b) p-type $Hf_{0.5}Zr_{0.5}CoSn_{0.25}Sb_{0.75}$ with a hole doping of $4.4 \times 10^{21}$ cm$^{-3}$, (c) Intrinsic $Hf_{0.625}Zr_{0.375}NiSn$, and (d) n-type $Hf_{0.625}Zr_{0.375}NiSn_{0.875}Sb_{0.125}$ with an electron doping of $2.2 \times 10^{21}$. The red dashed line indicates the Fermi level.

We find that the Fermi level is sensitive to the Brillouin zone sampling. Figure 2b shows the calculated electron concentration as a function of Fermi level with respect to the CBM for the Monkhorst-Pack shifted k-grids of 8x8x4, 24x24x12 and 36x36x18, indicating that using 36x36x18 k-sampling yields well converged values for the carrier concentration and Fermi level. Therefore, the Fermi level used in this work is based on these converged calculations. We find that the electron concentrations of $2.2 \times 10^{21}$ cm$^{-3}$ and $1.3 \times 10^{20}$ cm$^{-3}$ for the n-type HH correspond to the Fermi level being 346 meV and 40 meV above the CBM, respectively (See Figure 1d and Figure 2b), and the hole concentration of $4.4 \times 10^{21}$ cm$^{-3}$ for the p-type HH results in a Fermi level 211 meV below the VBM (See Figure 2a).

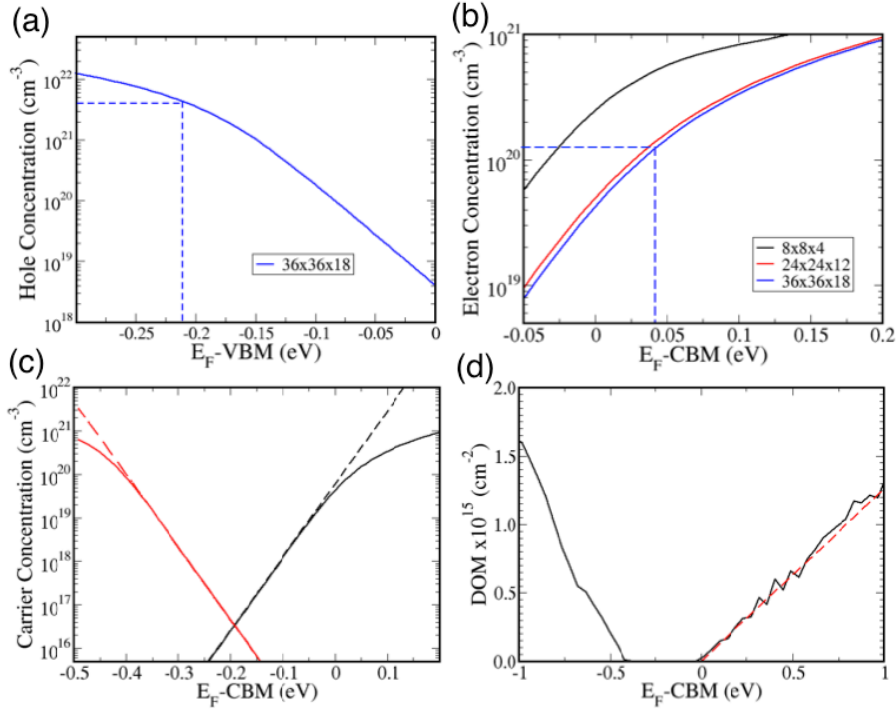

**Figure 2.** Calculated carrier concentration and density of modes (DOM) as a function of Fermi level. (a) Hole concentration for $Hf_{0.5}Zr_{0.5}CoSn_{0.25}Sb_{0.75}$ obtained with a k-grid of 36x36x18; (b) Electron concentration for $Hf_{0.625}Zr_{0.375}NiSn_{0.875}Sb_{0.25}$ obtained with different k-samplings (i.e. 8x8x4, 24x24x12 and 36x36x18). For the unconverged k-grids 8x8x4, 24x24x12, the meaning of CBM is that of the lowest energy of the conduction states sampled within these respective k-grids. The dashed blue lines in (a) and (b) represent the extracted values of Fermi level corresponding to the hole and electron concentrations used in this work. (c) DOS effective mass fitting for holes (red line) and electrons (black lines) for the n-type material $Hf_{0.625}Zr_{0.375}NiSn_{0.875}Sb_{0.25}$. (d) DOM effective mass fitting for the n-type material.

Having understood the characteristics of the doped bulk HH systems, we investigated the interfacial properties between the HHs and Ag. We chose Ag as the metal contact because of its high electrical and thermal conductivity and reliable performance over a wide range of temperatures including good oxidation resistance. Relatively low electrical resistivity has been reported for HH-Ag interfaces in the medium temperature range below 450 ºC[14]. We note that the linear expansion coefficient of Ag (2e-5 K$^{-1}$) matches very well the one of p-type HH[14], which can be beneficial for the stability of the contact at high temperatures as it reduces the possibility of microgaps formation at the interface.

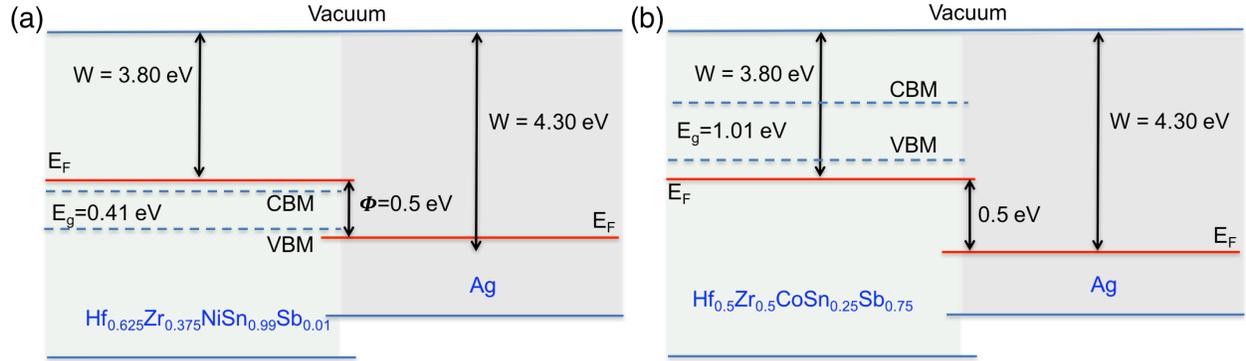

**Figure 3.** Schematic band alignment at the interface between HH alloys and Ag, based on the calculated work functions (W) of Ag and for (a) n-type ($Hf_{0.625}Zr_{0.375}NiSn_{0.99}Sb_{0.01}$) and (b) p-type ($Hf_{0.5}Zr_{0.5}CoSn_{0.25}Sb_{0.75}$). CBM and VBM represent the conduction band minimum and valence band maximum. $E_F$, $E_g$ and $\Phi$ are the Fermi level, bandgap and Schottky barrier, respectively.

We aim at establishing the fundamental limit of contact resistivity, and thus consider clean HH-Ag interfaces, neglecting oxidation or interface reactivity[14]. We first used the conventional Schottky-Mott rule to estimate the band bending in the HH/Ag systems. The work function values (W) for the doped HH and for Ag were obtained from the difference between the vacuum electrostatic potential ($V_{vacuum}$) and Fermi level ($E_F$) calculated using *ab initio* simulation with isolated slabs of the HH or Ag. The slab thickness was chosen large enough to restore the bulk properties. As shown in Figure 3a and 3b, the band alignments based on the calculated work functions suggest that the n-type HHs contacted with Ag form a Schottky contact with a barrier height of ~0.5 eV, but the p-type HHs and Ag form Ohmic contacts.

This "bare" band bending obtained from the Schottky-Mott rule is based on the bulk properties, and does not consider atomistic interfacial effects. Such effects are known to significantly influence the band alignment, for example by pinning the Fermi level. To assess the role of atomistic interfacial effects, we implemented a combination of macroscopic model and *ab initio* calculations. To study the influence of the interfacial structure on the band bending, we generated two interfacial structures for both the n-type materials (i.e. HfZrSn/Ag and Ni/Ag) and the p-type materials (i.e. HfZr/Ag and Ni/Ag). We find that the interfacial structures strongly depend on the chemical species at the interface between the HH and Ag. If the HH surfaces are terminated with HfZrSn for n-type or HfZr for p-type, the interfacial structure is quite ordered; whereas if the contacted HH surface is terminated with Ni, it becomes disordered (see Figs 4a and 4b) with significant relaxation taking place in both the HH and Ag sides of the interface. All generated interfacial structures have 1.2nmx1.2nm cross sections chosen based on the convergence test of the size effect on the disordered structures, leading to an interfacial strain of ~2%. We find that the disordered/ordered interfaces formed after optimization are intrinsic phenomena, which depend on the chemical species terminating the surfaces of the different materials, not on lattice mismatch. For the case of HH terminated with Ni, the disordered structure appears in about 1 nm range about the interface.

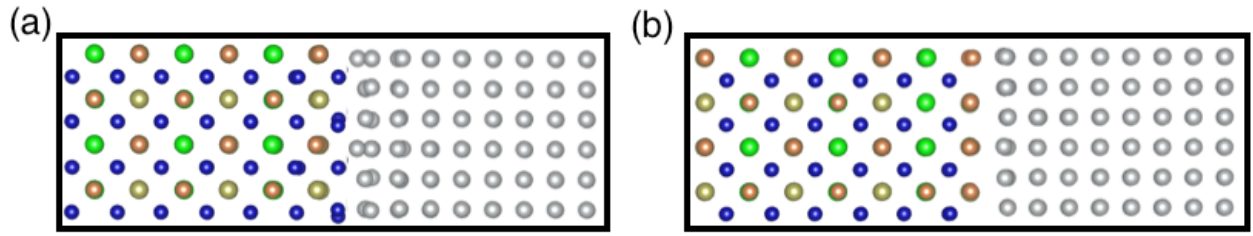

**Figure 4.** Atomic structure of HH/Ag interfaces optimized using *ab initio* calculations. (a) n-type HH terminated with Ni. Similar results are obtained for the p-type HH terminated with Co. (b) n-type HH terminated with HfZrSn. Similar results are obtained for the p-type HH terminated with HfSb and ZrSbSn.

Having obtained the atomic structures of the electrical contacts, we calculated the electrostatic potential and charge distribution at the interfaces. Figures 5a and 5b show the planar-averaged electrostatic potential difference (V) and charge density difference (Q) calculated along the transport direction perpendicular to the HH and Ag interface for two n-type systems and one p-type system. We see that the electrostatic potential close to the interface depends not only on the interfacial structure but also on doping, while Q is mainly dominated by the interfacial atomic structure. For the n-type doped HH, the electrostatic potential bends up close to the metal interface for both ordered and disordered interfacial structures. However, within 5 Å from the interface, V is significantly different between the ordered and disordered cases. On the other hand, Figure 3d shows that the charge density difference mostly appears within 5 Å for all three systems, with larger values for the ordered interfaces and smaller values for the disordered interface. These results indicate that the band bending is dominated by the doping at distances larger than 5 Å, but at distances smaller than 5 Å, it is controlled by the interfacial charges (see Fig 5b). The different values of charge density difference indicate that the ordered contact interface terminated with HfZr (or HfZrSn) induces a larger charge and tends to bend the bands down, while the disordered contact interface terminated with Ni leads to a smaller value of charge and tends to bend the bands up.

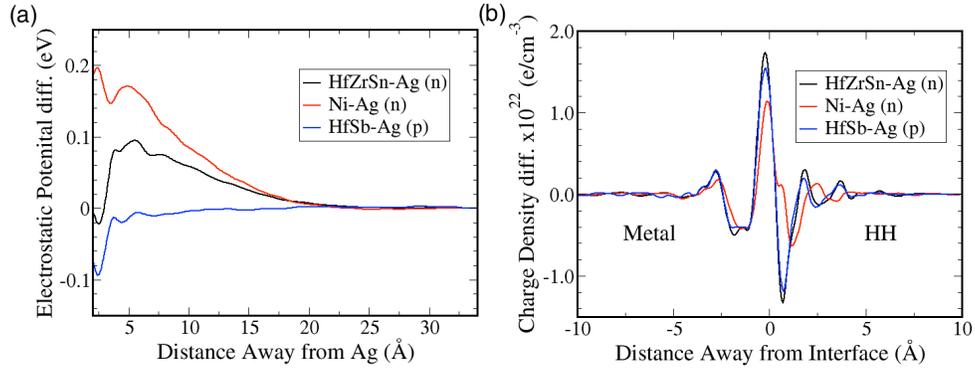

**Figure 5.** (a) Electrostatic potential difference in the HH from the *ab initio* calculations. (b) Charge density difference in both the HH and the metal as a function of distance away from the HH/metal interface. Black, red and blue are for the HH surface terminated with HfZrSn (n-type), Ni (n-type) and HfSb (p-type), respectively.

The interfacial atomistic effects play a critical role in determining the band alignment. In the case of the p-type HHs, the Schottky-Mott rule already predicts an ohmic contact, and this picture is maintained upon consideration of atomistic effects. However, for the n-type contacts the Schottky-Mott rule completely fails in quantitatively predicting the eventual band alignment. Indeed, as shown in Fig. 6a, the atomistic effects lead to a Schottky barrier of at most 0.1 eV, compared to the predicted value of ~0.45eV. The height of the Schottky barrier is larger for the disordered interface (0.1eV) compared to the ordered interface (0.05eV). A similar conclusion can be drawn from the *ab initio* calculated projected bandstructure near the interface (see fig. S1 in Supplemental Material ).

Since the interface between the p-type HH and Ag is ohmic, and for the n-type HH it has a Schottky barrier, the thermoelectric efficiency of a TE module composed of n-type and p-type HH in direct contact with Ag could be limited by the contact resistance of the n-type HH. Using macroscopic modeling with the *ab intio* input parameters, we obtained the contact resistivity for the n-type HH. We calculated the contact resistivity using the band bending at distances larger then 5 Å away from the interface for both ordered (i.e. HfZrSn/Ag) and disordered (i.e. Ni/Ag) interfacial structures. This is motivated by the fact that the projected density of states as a function of distance from the interface shows a large number of electronic states within 5 Å which might provide extra electron transport channels ( see Figs 6c and 6d as well as Fig S1 in Supplemental Material).

We used our macroscopic modeling approach to study the dependence of the contact resistivity on doping and temperature, assuming a clean interface could be maintained in the low to medium temperature range. As shown in Fig. 7, we see that doping can significantly reduce the

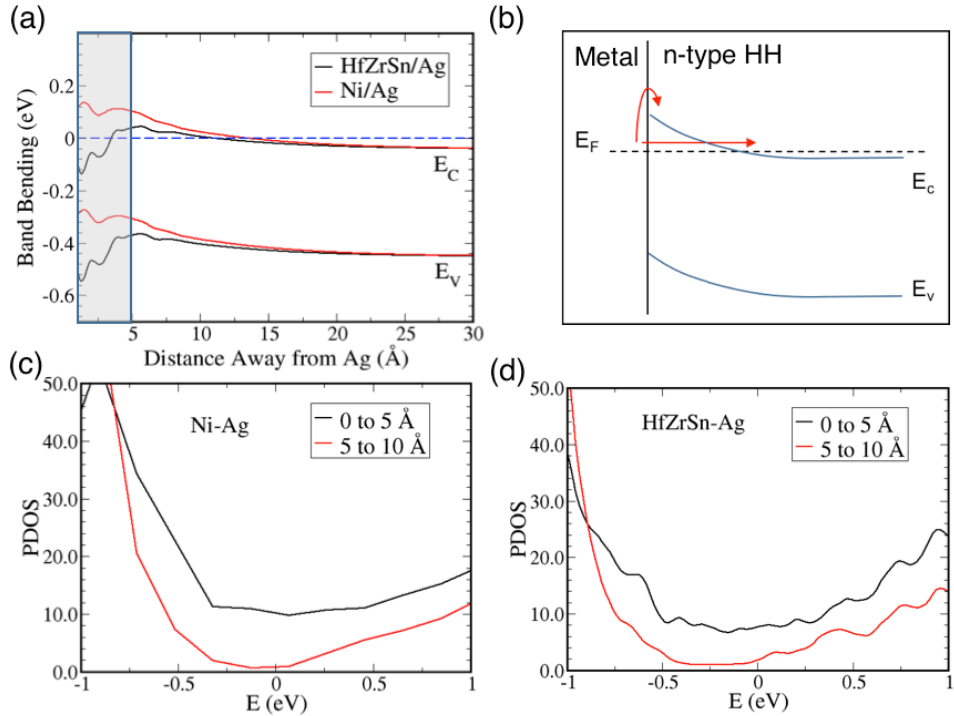

**Figure 6.** (a) Calculated interfacial band bending for the n-type HH with the HfZrSn/Ag (ordered) and Ni/Ag (disordered) interfaces using *ab initio* based modeling. $E_c$ and $E_v$ are the CBM and VBM, respectively. The blue dash line indicates the Fermi level. (b) Illustration of the band bending leading to thermionic emission and tunneling from the metal to n-type HH. The barrier width and height are obtained from macroscopic simulations benchmarked with *ab initio* simulations. Panels

(c) and (d) show the calculated projected density of states (PDOS) around the Fermi level (E=0 eV) for the n-type HH terminated with Ni and HfZrSn, contacted with Ag.

contact resistivity; this arises because the band-bending becomes sharper which favors tunneling across the Schottky barrier instead of thermionic emission over the barrier (see Fig. 6b). This is in agreement with well-known approaches widely used to optimize the interfacial transport properties of semiconductor-metal contacts[34-36]. The contributions from thermionic emission and tunneling are also apparent when considering the temperature dependence. Indeed, Fig. 7 shows that at low doping the contact resistivity varies by an order of magnitude with temperature since the electrons need to be thermally excited over the Schottky barrier. In contrast, the temperature dependence is weak at large doping due to the dominance of direct tunneling at the Fermi level.

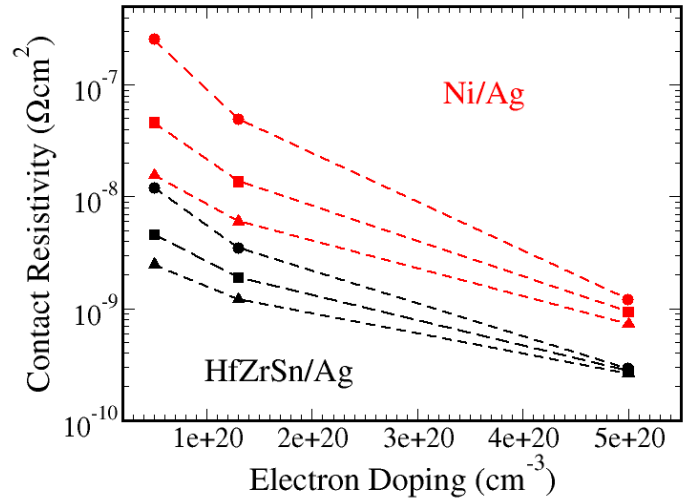

**Figure 7.** Calculated contact resistivity as a function of doping concentration for the n-type HH (black for HfZrSn/Ag and red for Ni/Ag) at different temperatures: circle, 300K; square, 400K; triangle, 500K. The dash line is a guide for the eye.

## IV. Conclusion

Half-Heusler(HH) alloys have been identified as promising thermoelectric (TE) materials in the medium and high temperature range due to their optimal high temperature TE properties.

To harness HH materials for TE devices, it is important to realize metal contacts with low electrical contact resistivity. Although several metal electrodes contacted with HH have been explored experimentally, there are no theoretical studies of metal contacts to HH alloys. We present a theoretical study of the fundamental limits of contact resistivity to HH alloys. By using a combination of *ab initio* calculations and macroscopic modeling, we investigate the detailed interfacial structure and electronic properties, as well as their impact on the contact resistivity of an n-doped HH alloy ($Hf_{0.75}Zr_{0.25}NiSn$) and a p-doped HH alloy ($Hf_{0.5}Zr_{0.5}CoSb$) contacted with the metal Ag as a function of doping concentration in the low to medium temperature range. We find that the surface termination of the HH plays an important role in determining the properties of the interface. In cases where the HH termination contains Ni, we find a disordered interfacial structure, whereas terminations with HfZrSn or HfZr give ordered interfaces. The detailed atomistic electronic structure of the interface is also essential in determining the band alignment at the interface: for n-type contacts, the simple Schottky-Mott rule predicts a large Schottly barrier, however we find that interfacial effects significantly reduce the Schottky barrier, rendering the contact almost ohmic.

Our theoretical study is based on ideal interfacial structures realizable in principle at low temperature. More complex scenarios are possible in devices operated at elevated temperatures, including the formation of new phases at the interface due to the chemical reactivity between HH and the metal. However, we note that the predicted Schottky barrier for n-type HH interface, and ohmic contact for p-type HH interface with Ag are coincident with the experimental observations[14], where the contact resistivity of p-type HH with Ag is not very sensitive to temperature (T), while for the n-type HH it strongly depends on T. In particular, at low T the contact resistivity for the n-type HH interface is much larger than that of the p-type HH interface. Nevertheless, the effect of interfacial phases formed at high temperature on the contact resistivity needs to be further investigated.

Finally, our theoretical work serves to establish the fundamental understanding of metal contacts to HH via atomistic insight into the behavior of the band bending near the interface including the impact of the interfacial structure and chemical composition. Our calculations provide a lower bound of contact resistivity of the interface of HH and metal Ag, which can be used as a reference for future studies.

## ACKNOWLEDGMENTS

This research was developed with funding from the Defense Advanced Research Projects Agency (DARPA) and supported by the DARPA MATRIX program. The views, opinions, and/or findings contained in this article/presentation are those of the author(s)/presenter(s) and should not be interpreted as representing the official views or policies of the Department of Defense or the U.S. Government. Sandia National Laboratories is a multimission laboratory managed and operated by National Technology and Engineering Solutions of Sandia, LLC., a wholly owned subsidiary of Honeywell International, Inc., for the U.S. Department of Energy's National Nuclear Security Administration under contract DE-NA-0003525.

# Supplemental Material for 'Atomistic Study of the Electronic Contact Resistivity Between the Half-Heusler Alloys (HfCoSb, HfZrCoSb, HfZrNiSn) and the Metal Ag'

Yuping He, François Léonard and Catalin D. Spataru*

*Sandia National Laboratories, Livermore, California, 94551, USA*
*E-mail: cdspata@sandia.gov

Fig. S1 shows the *ab initio* calculated projected electronic bandstructure near the interface of n-type HH and the metal Ag. The conduction band minimum (CBM) inside the HH can be identified for distances away from interface larger than ~ 1nm. The relative position of the CBM w.r.t. the Fermi level in the metal (zero energy level on the y-axis) as function of distance is in good agreement with the calculated interfacial band bending shown in fig. 6(a) of the main text.

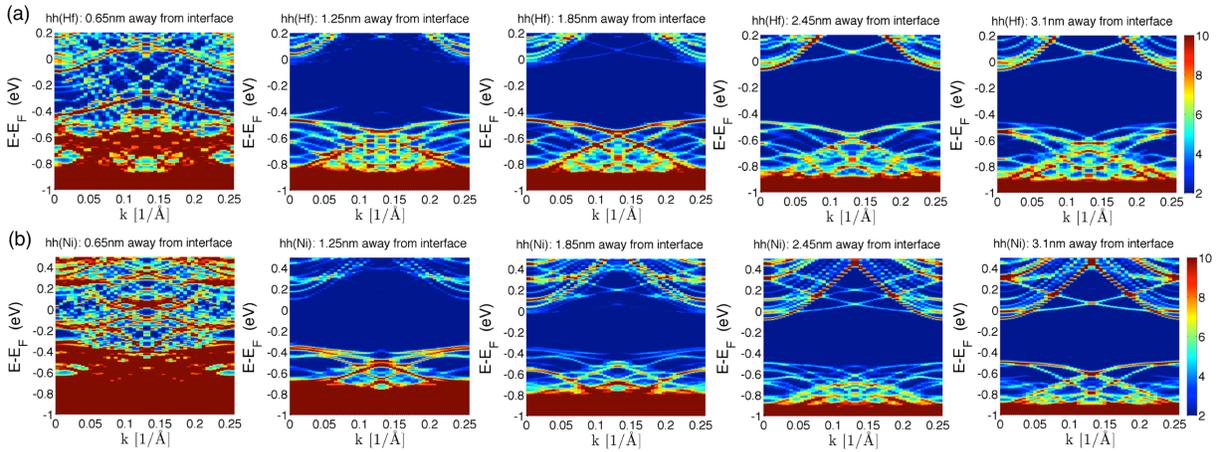

Figure S1. Calculated projected electronic bandstructure of the n-type HH/Ag interface. The spectral function projected inside the HH on a slab of width ~0.6 nm is shown for several distances $d_i$ between the interface and the side of the slab farthest from the interface (i.e. $d_i = 0.65$ nm, 1.25 nm, 1.85 nm, 2.45 nm and 3.1 nm). (a) for the HH surface terminated with HfZrSn [hh(Hf)] and (b) for the HH surface terminated with Ni [hh(Ni)], using the definition for the projected spectral function: $A_j(\vec{k}, E) = \sum_{n,i}^{i \in d_i} w_{n\vec{k}}^i \delta(E - \varepsilon_{n\vec{k}})$, where $w_{n\vec{k}}^i$ is the site-projected character of the wave function of an electron characterized by band index n and crystal momentum $\vec{k}$, i is the index of the atomic site. The δ function centered on the electron energy $\varepsilon_{n\vec{k}}$ is broadened by a Lorentzian function of width equal to 10 meV.